\documentclass[reprint, amsmath,amssymb,pre]{revtex4-2}

\usepackage{graphicx}
\usepackage{dcolumn}
\usepackage{bm}

\usepackage{xcolor}

\begin{document}

\preprint{APS/123-QED}

\title{Effects of particle angularity on granular self-organisation}

\author{Dominik Krengel}
 \email{dominik.krengel@kaiyodai.ac.jp}
\affiliation{
 Department of Marine Resources and Energy, \\
 Tokyo University of Marine Science and Technology,\\
 4-5-7, Konan, Minato, 108-8477, Tokyo, Japan
}%
\author{Haoran Jiang}
\email{jiangh@kajima.com}
\affiliation{
 Civil Engineering Design Division, Kajima Corporation,\\
 3-8-1, Motoakasaka, Minato-ku, Tokyo, 107-8477, Japan
}%
\author{Takashi Matsushima}%
 \email{tmatsu@kz.tsukuba.ac.jp}
\affiliation{
 Department of Engineering Mechanics and Energy, University of Tsukuba,\\ 
 1-1-1, Tennodai, Tsukuba, 305-8573, Japan
}%
\author{Raphael Blumenfeld}
\email{rbb11@cam.ac.uk}
\affiliation{
 Gonville \& Caius College, University of Cambridge, UK
}%
\altaffiliation{Also at Imperial College London, London SW7 2AZ, UK \\
}%

\date{\today}

\begin{abstract}
Recent studies of two-dimensional poly-disperse disc systems revealed a coordinated self-organisation of cell stresses and shapes, with certain distributions collapsing onto a master form for many processes, size distributions, friction coefficients, and cell orders. Here we examine the effects of particle angularity on the indicators of self-organisation, using simulations of bi-disperse regular $N$-polygons and varying $N$ systematically. 
We find that: the strong correlation between local cell stresses and orientations, as well as the collapses of the conditional distributions of scaled cell stress ratios to a master Weibull form for all cell orders $k$, are independent of angularity and friction coefficient. In contrast, increasing angularity makes the collapses of the conditional distributions sensitive to changes in the friction coefficient. 

\end{abstract}

\maketitle

\section{Introduction}\label{sec:intro}
Understanding behaviour of granular matter is essential in many fields of research and technology. Macroscale models of stress transmission in such media involve upscaled constitutive properties that often presume independence of the stress field, which enables explicit solutions. The constitutive properties are sensitive to the local particle-scale structural characteristics, but these were shown recently to be strongly correlated with the local stress due to a cooperative local stress-structure self-organisation~\cite{Matsushima2017,Matsushima2021,Jiang2022}. Significantly, this calls into question models involving linear stress field equations.
While relation between macroscopic properties and particle-scale characteristics, such as force-chains, have been studied extensively, advances on effects of dynamics on the settled structure have been limited~\cite{Nguyen2009,Zhu2016,Zhu2016a}. Recent studies of planar (2D) disc packings have shown that the basic organisable structural elements, cells, which are the smallest voids surrounded by particles in contact, self-organise during slow dynamics in ways that transcend particular system and process details~\cite{Matsushima2014,Matsushima2017, Matsushima2021,Sun2020,Jiang2022}, including remarkable steady-state features~\cite{Wanjura2020,Sun2021,Myhill2023,Wanjura2024}.

Self-organisation (SO) is a ubiquitous phenomenon, observed in many aspects of life, from natural phenomena to social behaviour~\cite{SOR1,SOR2,SOR3,SOR4,SOR5}. Its emergence depends on a wide range of variables, some of which are continuous. This makes a general theory difficult, although some attempts appeared recently~\cite{SOR4,Chvykov2021,Blumenfeld2025}. In the context of granular dynamics, many signatures of SO have been observed in the last few years and one question is how these are affected by particle angularity. Not only are non-spherical particles more ubiquitous in nature~\cite{Ulusoy2023} but they also restrict mobility~\cite{Flemmer1993,Cleary2008,Wang2010,Kozlowski2022} and affect the stability of granular aggregates~\cite{Wouterse2007,Athanassiadis2014,Krengel2015,Altuhafi2016}. Owing to the involved analytical and numerical treatments of such shapes, a limited body of  numerical work exists on micro- and meso-scale such systems, e.g.~\cite{NouguierLehon2005,Carlevaro2011,Azema2012,Miao2017,Krengel2023,Jiang2024,Hua2022}. In contrast, the effects of inter-particle friction has received more attention~\cite{Yang2012,Osipov2016,Sun2020,Wei2022,Nie2023,Lai2023}, although primarily in systems of discs and spheres. However, the effects of shape and friction on the behaviour of particles are not independent of each other. While these combined effects on bulk properties have received some attention~\cite{Binaree2020, Binaree2021, Krengel2025}, the effects on the micro-structure are not fully understood. In particular, studies of effects on cell characteristics in 2D have been limited so far to squares with rounded corners~\cite{Jiang2024a,Jiang_PhD}.

A way to explore effects of particle angularity in 2D, is by observing packings of $N$-sided regular polygonal particles, studying the effects of modifying $N$, and checking consistency in the discs limit, $N\to\infty$. In particular, we investigate systematically the interplay between angularity and friction and which indicators of SO~\cite{Krengel2023,Jiang2021,Jiang2022} survive the increasing angularity.

\section{Numerical Simulations}\label{sec:setup}
\subsection{Simulation setup}
\begin{figure}[h]
 \centering
 \includegraphics[width=\columnwidth]{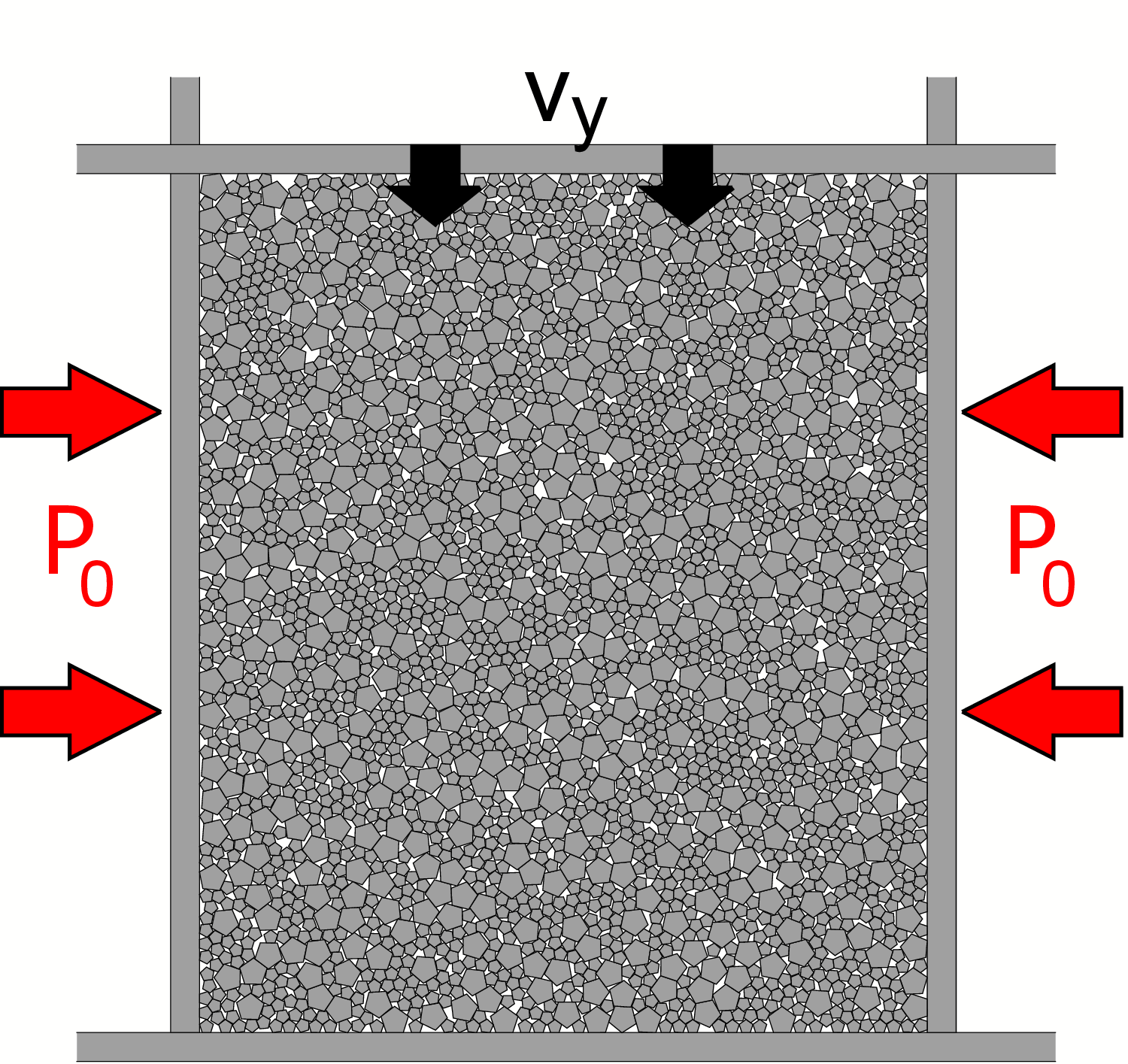}
 \caption{The biaxial shear setup and boundary conditions. The top wall is lowered down at a constant speed $v_y$ and the side walls are free to move to maintain a constant confining pressure $P_0$.}
\label{fig:shear_setup}
\end{figure}

We performed biaxial compression tests on $N$-edged regular polygon systems, with $N=5,6,13$, and $64$. $N=5$  was chosen because it is the most angular, non-space filling shape. $N=64$ because the kinematics of such systems are hardly different from discs and $N=13$ was chosen to provide intermediate data between the former two. $N=6$ was chosen as a reference point because structures of granular media often display hexagonal symmetry in 2D. $N=3$ and $4$ were omitted because these readily organize into polycrystalline structures~\cite{Wang2015}, sometimes also with bidisperse size distributions.
We used the hard-particle soft-contact Discrete Element Method for our simulations~\cite{Matuttis2014,Krengel2023}. In each simulation, 2200-2750 frictionless particles of equal shape were deposited under gravity initially to generate dense packings. We used the same initial domain for all simulations, resulting in less angular particles starting from denser packings. To minimize crystallization, we used an 80\%-20\% small-to-large mixture of sizes in each run, with circum-radii $r= 2.5$ and $5$mm, respectively. Further simulation parameters are given in table~\ref{tab:matpar}.
Once generated, gravity and sidewall friction were disabled to create isotropic conditions. The systems were compressed uniaxially by adjusting the positions of the left, right, and top walls until the pressure on all walls reached $P_0=20$kPa. Energy was then allowed to dissipate until pressure fluctuations decayed to less than $10^{-14}$ of the constant pressure. At this 'initial' state, intergranular friction, $\mu$, was switched on and the packing was compressed biaxially by moving the top wall down at constant speed $v_y$ and adjusting the left and right wall to maintain constant $P_0$, as sketched in Fig.~\ref{fig:shear_setup}. $P_0$ was kept constant, using a servo-mechanism~\cite{Cundall1988}.
For each value of $N$, we simulated systems with five different friction coefficients, $\mu=0.01, 0.1, 0.5, 1.0, 10.0$, each starting from a different initial random configuration. In total, 20 different combinations of $N$ and $\mu$ were simulated, each five times, for good statistics.
The pure shear, generated by the biaxial compression, continued well after the global stress plateaued out. The system's height at time $t$, $h(t)$, decreases relative to  $h(t=0)$ and the strain is defined as
\begin{equation}
 \epsilon =  (h(t=0)-h(t))/h(t=0) \ .
\end{equation}
In all our runs, all measurable quantities reached a steady state by the time $\epsilon$ reached $8\%$. Therefore, all steady-state data were collected between strains of $8$\% and $16$\%.

\begin{table}[t!]
 \centering
 \caption{Numerical parameters for the simulations.}
 \label{tab:matpar}
\begin{tabular}{ll}
 \hline
  Num. particles                         & $\sim$ 2200-2750\\
	Mixing ratio                           & 80\% small \\
	                                       & 20\% large, \\
  Circum-radii                           & 2.5 \& 5.0 [mm]\\
	Corner numbers $N$                     & $[5,6,13,64]$\\
  Friction coefficient $\mu$             & $[0.01, 0.10, 0.50,$\\ 
																				 & $ 1.00, 10.0]$\\
  Young's modulus $Y$                    & $10^9$ [N/m]\\
  Density $\varrho$                      & $2830$ [kg/m$^2$]\\
  Confining pressure $P_0$               & $20$ [kPa]\\
  Compression velocity $v_{y}$           & $10$ [mm/s]\\
	Ratio $Y/P_0$                   & $5\cdot 10^{5}$\\
  Timestep $\Delta t$                    & $2\cdot 10^{-6}$ [s]\\
 \end{tabular}
\end{table}

\subsection{Cells and cell stresses}
\begin{figure}[h]
 \centering
 \includegraphics[width=\columnwidth]{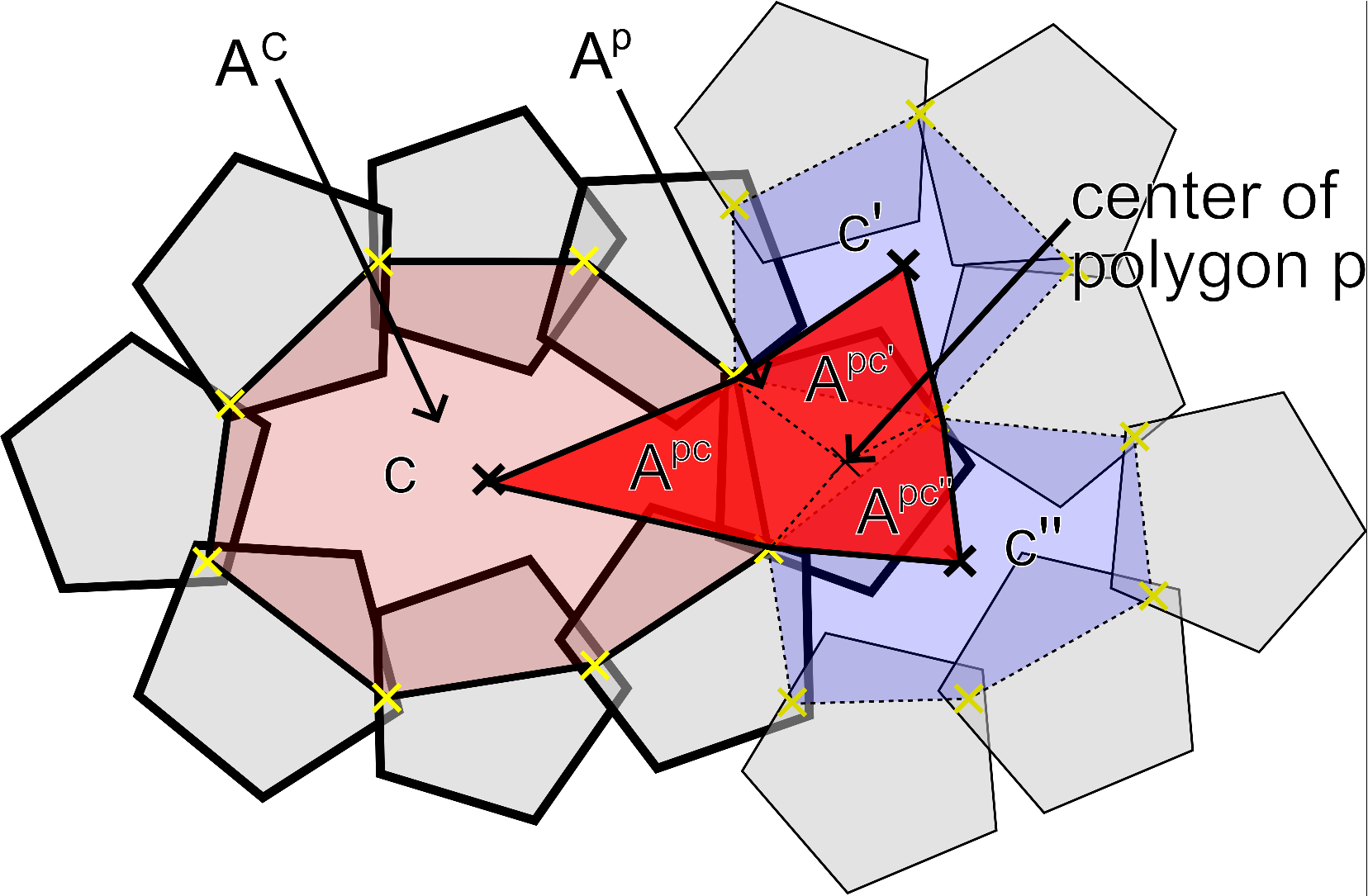}
 \caption{An example of a cell of order $k=8$: The cell is represented as a polygon, whose vertices are the centroids of the overlapping areas of its surrounding particles. $A_{pc},A_{pc'}$, and $A_{pc"}$ (in red) are the areas of the quadrons of $p$ extending into its surrounding cells, respectively, $c, c'$, and $c"$, and they represent the area associated with polygon $p$. The cell area is the sum of the quadrons of the particles extending into it. }
 \label{fig:celldefinition}
\end{figure}

In DEM, particles can penetrate slightly into one another and a contact between two particles is defined as the centroid of their overlapping areas.  
A cell is then defined as the polygon whose vertices are the contacts between the particles surrounding it, e.g., the red-shaded area in Fig.~\ref{fig:celldefinition}.  
The cells in our simulations were generically convex, although this need not be the general case. The order of a cell, $k$, is defined as the number of particles surrounding it, $k\geq3$.  Our analysis is limited to $k\leq 8$ because of the relatively poorer statistics and short life of higher-order cells

We define a cell stress $\bm{\sigma}_c$ as follows. Starting from the stresses of the particles $p$ that surround cell $c$, 
\begin{equation}
 \bm{\sigma}_p=\frac{1}{A_p}\sum_{p'} \bm{l}_{pp'} \otimes \bm{f}_{p'p} \ ,
\end{equation}
with $p'$ the particles in contact with $p$, $\bm{l}_{p'p}$ vectors connecting the centroid of $p$ with the contact with $p'$, and $\bm{f}_{p'p}$ the force exerted by $p'$ on $p$.
The area associated with particle $p$, $A_p$, is the sum of the areas of its quadrons~\cite{Blumenfeld2004,Blumenfeld2006}: $A_{pc},A_{pc'},A_{pc"}$, which extend to its surrounding cells, respectively, $c,c',c"$, as shown in Fig.~\ref{fig:celldefinition}. In terms of these, the cell stress is 
\begin{equation}
 \bm{\sigma}_c = \frac{\sum_{p\in c}A_{pc}\bm{\sigma}_p}{\sum_{p\in c}{A_{pc}}},
\end{equation}
with $p\in c$ denoting the particles surrounding cell $c$.

\subsection{Visualization of cells and forces}
\begin{figure*}[t!]
 \centering
 \includegraphics[width=0.575\textwidth]{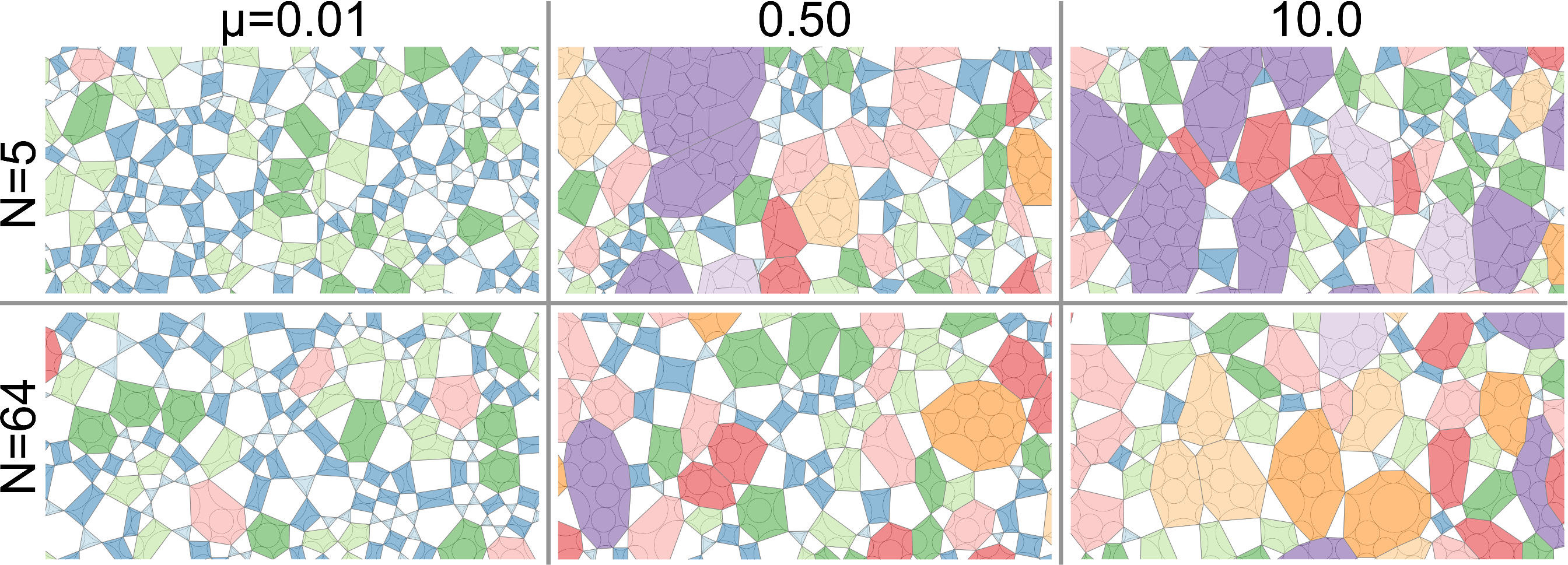}%
  \hfill
 \includegraphics[width=0.38\textwidth]{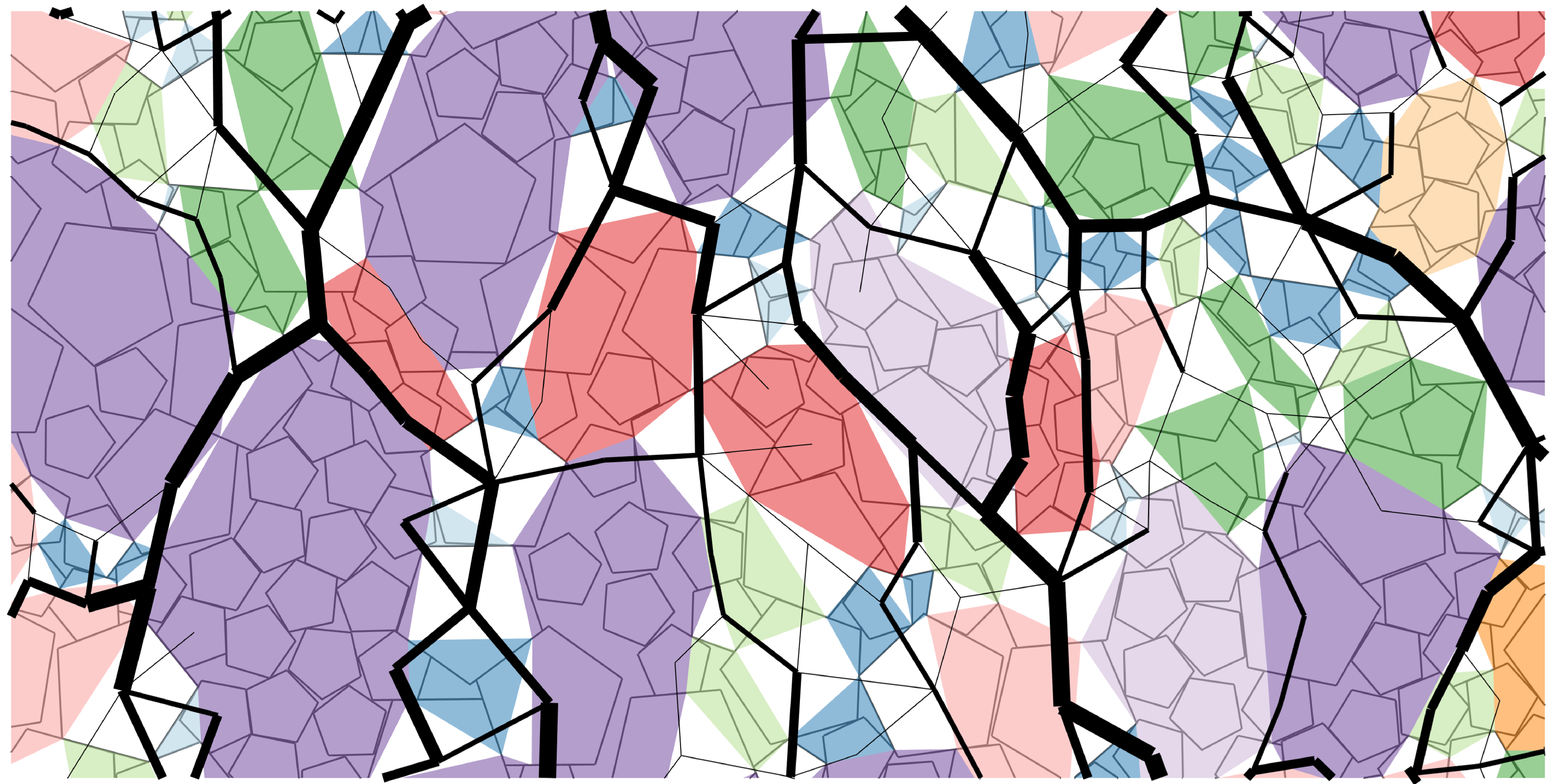}%
 \caption{(Left) Close-up view of steady-state cell structures of most and least angular particles when $\mu=0.01, 0.50$, and $10.0$. Cells are coloured according to their order. The fraction of high-order cells increases both with angularity and friction. These cells contain more rattlers, as evidenced by the superimposed force chains (black lines) in the right panel for $N=5,\mu=10.0$.}
 \label{fig:cell_visualisation}%
\end{figure*}

In Fig.~\ref{fig:cell_visualisation}, we show in the left panel examples of cells for several combinations of $N$ and $\mu$. When $\mu = 10.0$, angular particles readily form cells with $k>8$. The particles shown inside cells are rattlers, which are identified by visualizing the force chains network, outlined by the black lines, as illustrated in the right panel. 

\section{Results}\label{sec:results}
\begin{figure}[h]
\includegraphics[width=\columnwidth]{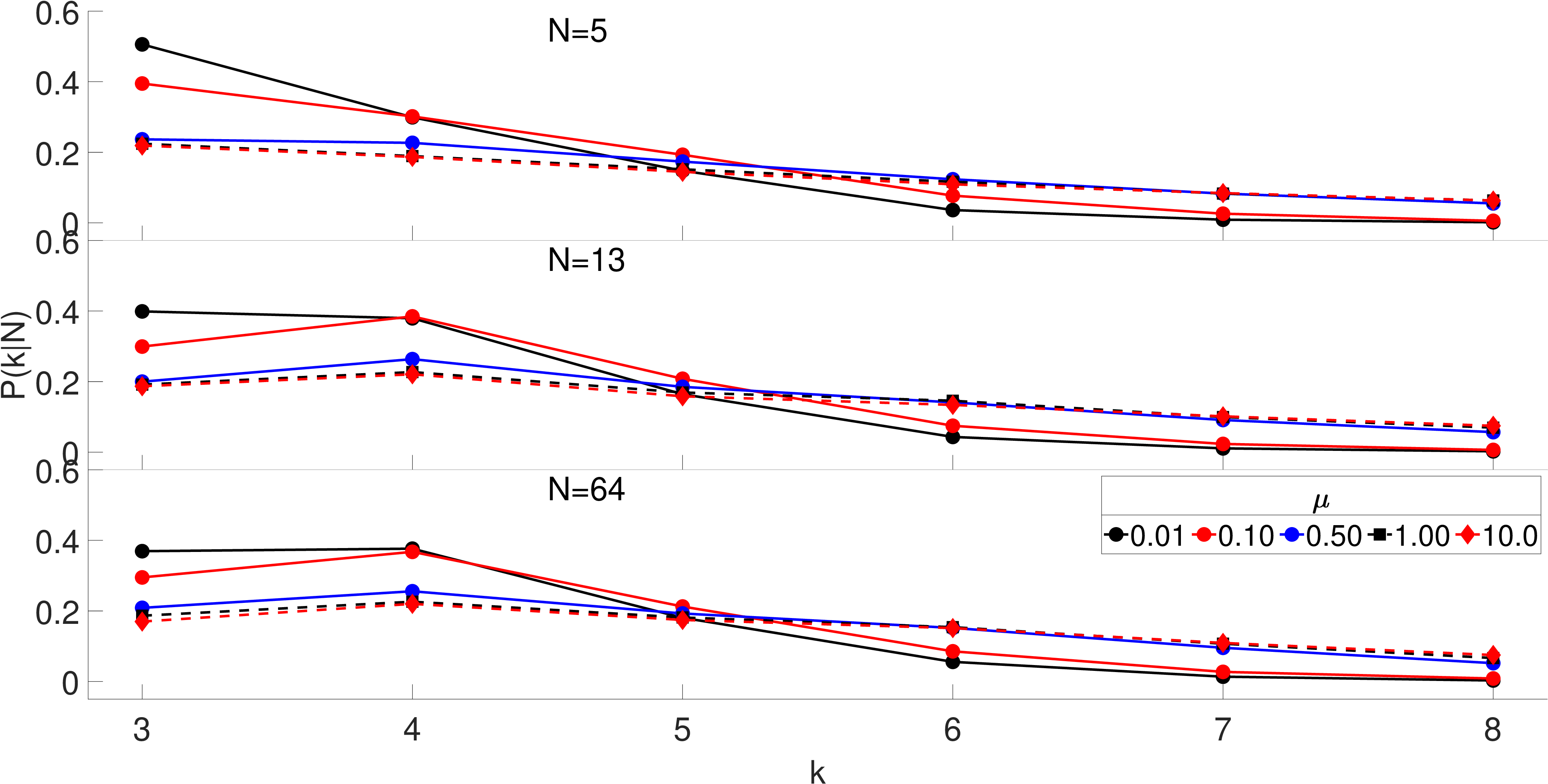}%
\caption{The steady-state conditional probabilities $P(k|N)$ of the cell orders $k$ for different angularities.}
\label{fig:cellorderdistrib}%
\end{figure}

We find that the conditional cell order probabilities, $P(k|N)$, depend on on both the angularity and on friction (Fig.~\ref{fig:cellorderdistrib}) -- the more angular the particles the more sensitive they are to $\mu$. Interestingly, while cells of order $3$ are the most frequent for $N\leq13$, cells of order $4$ are the most frequent when $N$ increases. As expected, increasing $\mu$ also increases the occurrence probabilities of high-order cells, but up to cell order $5$, $P(k\leq5|N)$ decreases with $\mu$ for all $N$. Reducing $N$ increases the proportion of order-$3$ cells and results in more compact structures. 

\begin{figure}[h]
 \centering
 \includegraphics[width=\columnwidth]{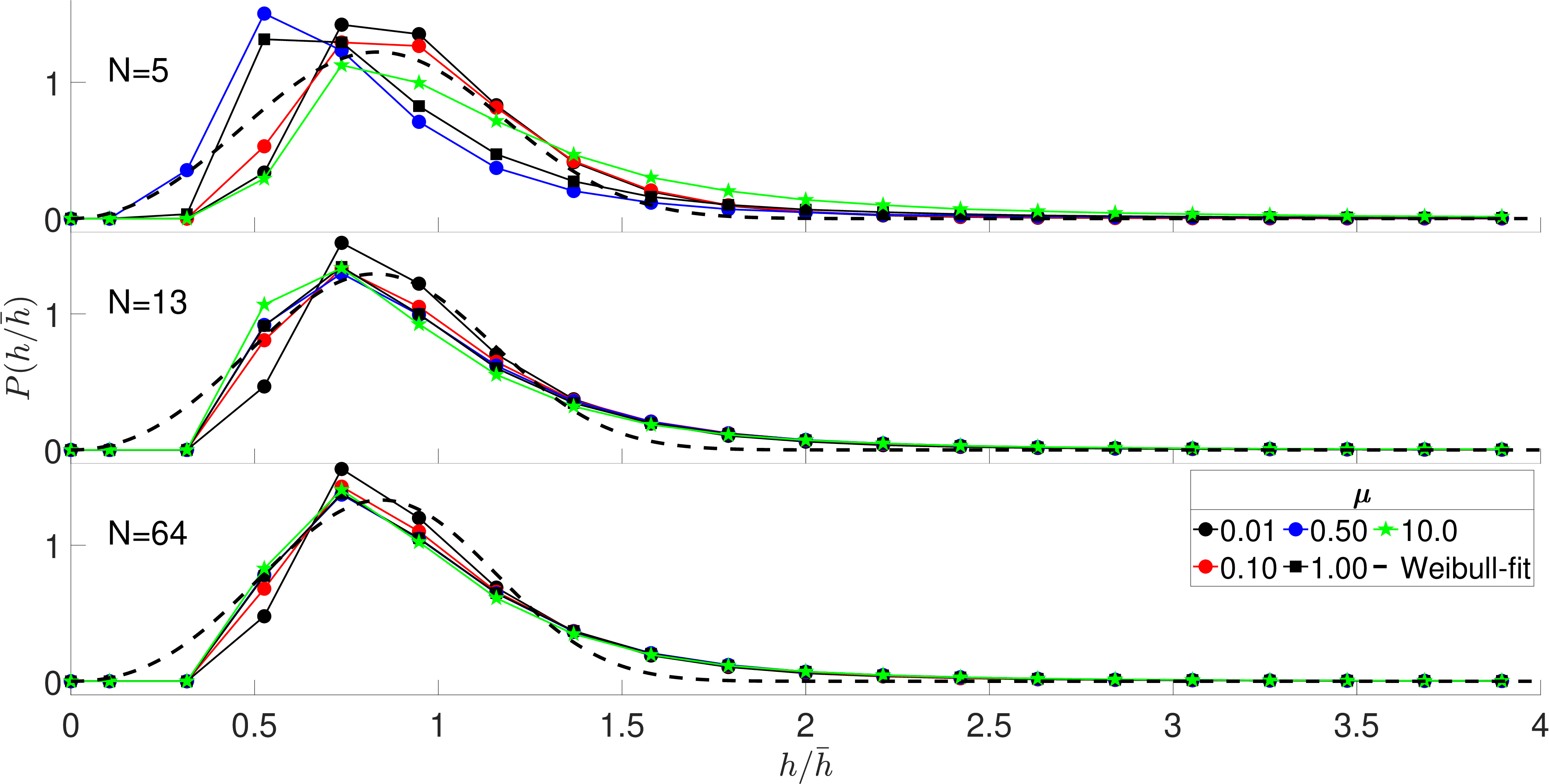}
 \caption{The collapsed form of $P(\hat{h})$ for different values of $\mu$ and $N$. The dashed lines represent Weibull-distribution fits for the different values of $\mu$.}
 \label{fig:cellstressratio}%
\end{figure}

\begin{figure}[h]
 \centering
 \includegraphics[width=\columnwidth]{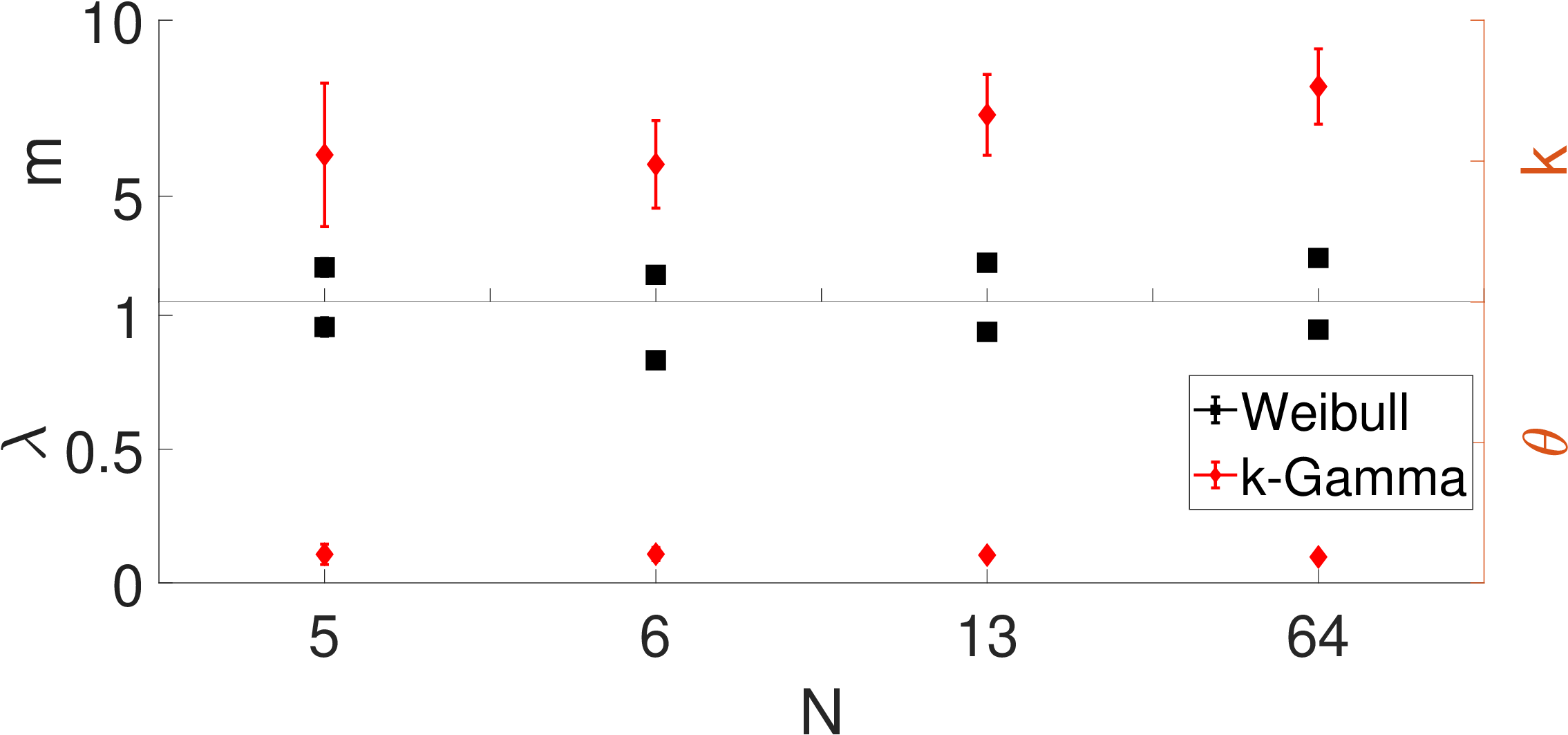}%
 \caption{$N$-dependence of the parameters of the Weibull distribution fit to the cell probability distributions in Fig.\,4 in the main text. For reference we also plot the parameters of a k-Gamma distribution to the same data. The error for the Weibull parameters is too small to be seen at this scale.}
 \label{fig:cellstressratio_fit}%
\end{figure}

Denoting the principal cells stresses of the cell stress $\sigma_{c1}$ and $\sigma_{c2} < \sigma_{c1}$, we consider the ratio $h=\left(\sigma_{c1} - \sigma_{c2}\right) /\left(\sigma_{c1} + \sigma_{c2}\right)$, which is a measure of cell stability.
The conditional probability density function (PDF) of $h$, $P(h|k)$, was found in~\cite{Jiang2022} to collapse to a Weibull distribution in disc systems for all $k$ and $\mu$, when scaled by the mean $\bar{h}_k$, $P\left(\hat{h}\equiv{h}/{\bar{h}_k}|k\right)=P\left(\hat{h}\right)$ $\forall k$. They also developed a model to explain the emergence of this distribution. The situation is similar in our systems, although not as ideal. As the mean of $h$ over cells depends on both $k$ and $\mu$, we first scaled the PDFs by $\bar{h}_k(\mu)$ and obtained the forms shown in Fig.~\ref{fig:cellstressratio}. While these collapse relatively well for $N=13,64$, the collapse deteriorates somewhat for the very angular polygons, $N=5$. Nevertheless, these collapses are still described well by a Weibull form (we demonstrate in Fig.~\ref{fig:cellstressratio_fit} that this fit is better than the k-Gamma PDF).
We then averaged over all values of $\mu$ to obtain a PDF of $h$ given $k$ and scaled all these by their mean, $\bar{h}_k$, These also collapsed to the Weibull form 
\begin{equation}
 P\left(\hat{h}_k\right) = \frac{m}{\lambda}\left(\frac{\hat{h}_k}{\lambda}\right)^{m-1} e^{-(\hat{h}_k/\lambda)^{m}} \ ,
\end{equation}
with $m$ and $\lambda$ hardly depending on $N$, $m = 3.1\pm0.1$ and $\lambda = 0.949\pm0.002$ (Fig.~\ref{fig:cellstressratio_fit}).  The larger relative error in $m$ compared to that of $\lambda$, suggests that $m$ is more sensitive to $\mu$ than $\lambda$.
For comparison, we tried to fit the PDFs $P\left(\hat{h}\right)$ with k-Gamma forms, 
\begin{equation}
 P\left(\hat{h}\right) = \frac{\hat{h}^{k-1} e^{-(\hat{h}/\theta)}}{\theta^{k}\Gamma(k)} \ ,
\end{equation}
and found that these fits are worse than the Weibull one, with the parameters no longer independent of $N$ (Fig.~\ref{fig:cellstressratio_fit}).


\begin{figure}[h]
 \centering
 \includegraphics[width=\columnwidth]{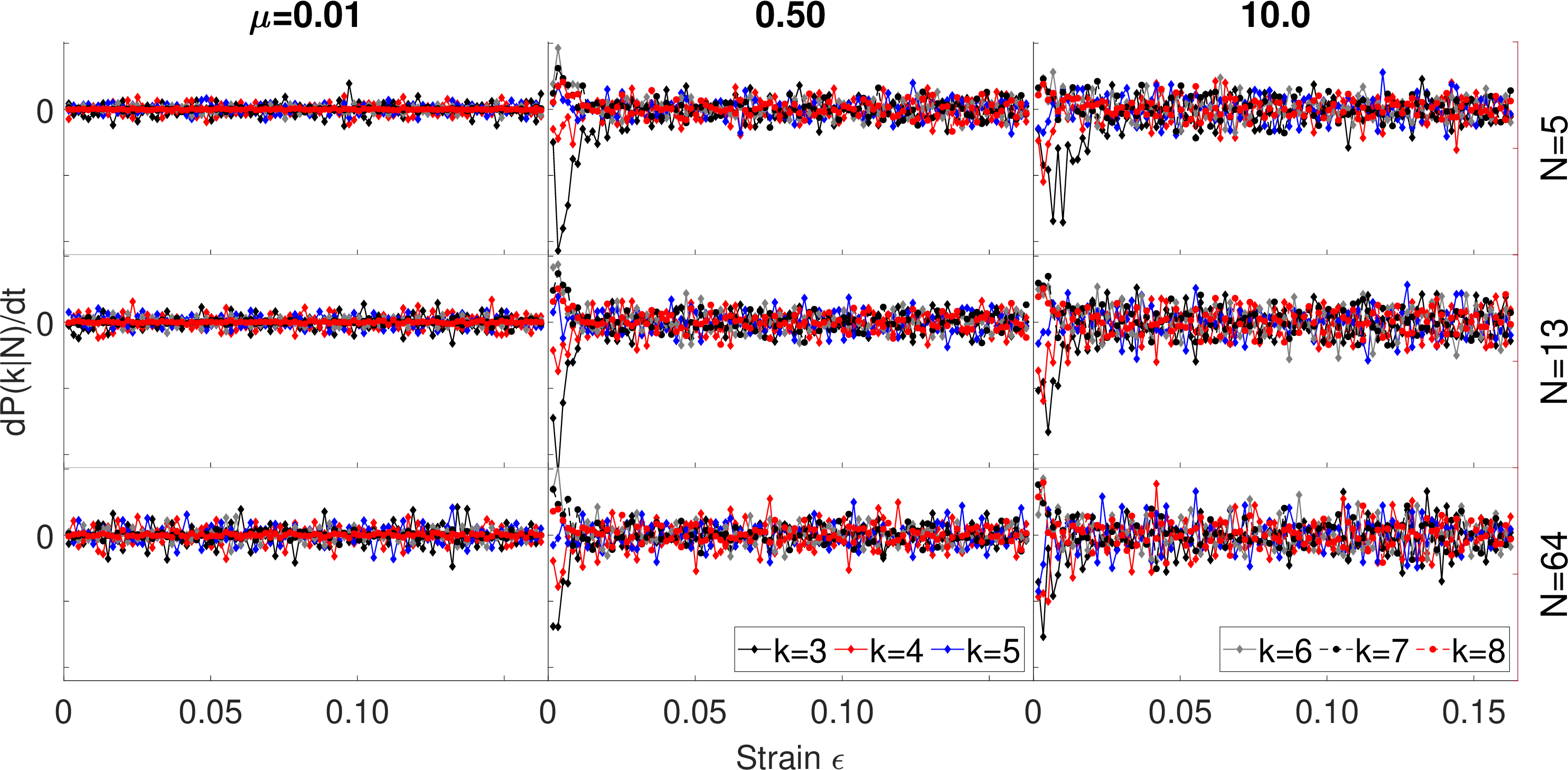}%
 \caption{The rate of change of the cell conditional probabilities, $dP(k|N)/dt$, with strain for different combinations of friction $\mu$ and angularity $N$.}
\label{fig:cellorderevolution}
\end{figure}

Fig.~\ref{fig:cellorderevolution} illustrates the convergence of the cell conditional probability distributions to their steady-state values, $\mathrm{d} P(k|N,t)/\mathrm{d}t\to0$, for different values of $N$ and $\mu$. The fluctuations around the steady state increase somewhat with $\mu$, which is expected to be a general feature. In contrast, the difference between the initial convergence of the very low-friction systems ($\mu=0.01$) and the other systems is attributed to the initial system preparation and the particular procedure we used to generate the steady state.

\begin{figure}[h]
 \includegraphics[width=\columnwidth]{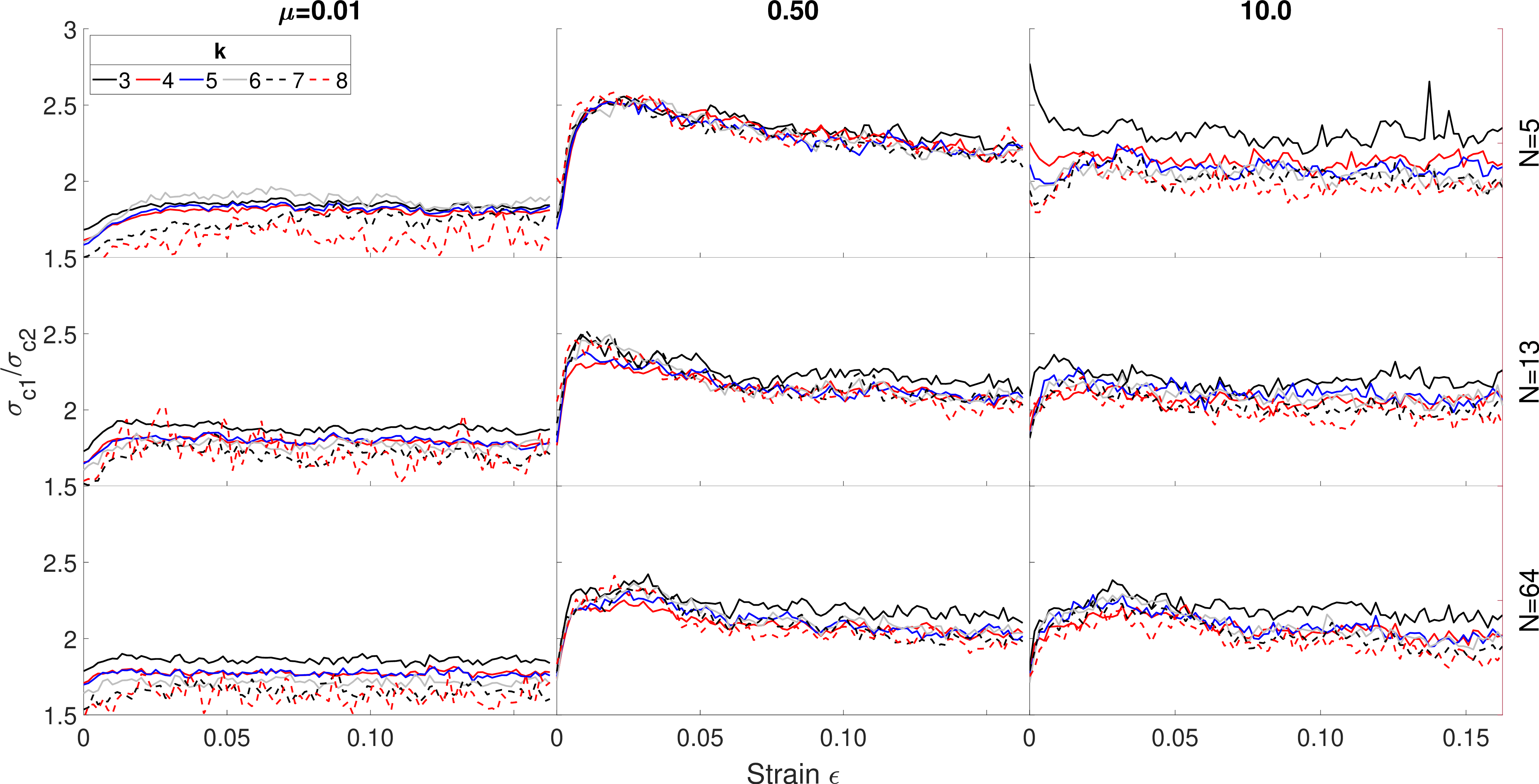}
 \caption{Evolution of the mean cell stress ratio with strain for each order of cell and different combinations of $\mu$ and $N$.}%
 \label{fig:cellstress_evolution}%
\end{figure}

In Fig.~\ref{fig:cellstress_evolution} we plot the mean cell stress ratio for each $k$, $N$, and $\mu$. 
While the initial evolution of the cell stress-ratio, $\sigma_{c1}/\sigma_{c2}$, is qualitatively different for the different combinations of $N$ and $\mu$, the steady-state values were very similar. 
The exception from the above is the most angular particles assembly, $N=5$, which we discuss in detail below and in the concluding section. Nevertheless, the convergence in almost all cases to a very similar steady-state stress ratio, observed here for the first time, supports further the idea of cooperative stress-structure self-organisation~\cite{Jiang2022}.

\begin{figure}[h]
 \includegraphics[width=\columnwidth]{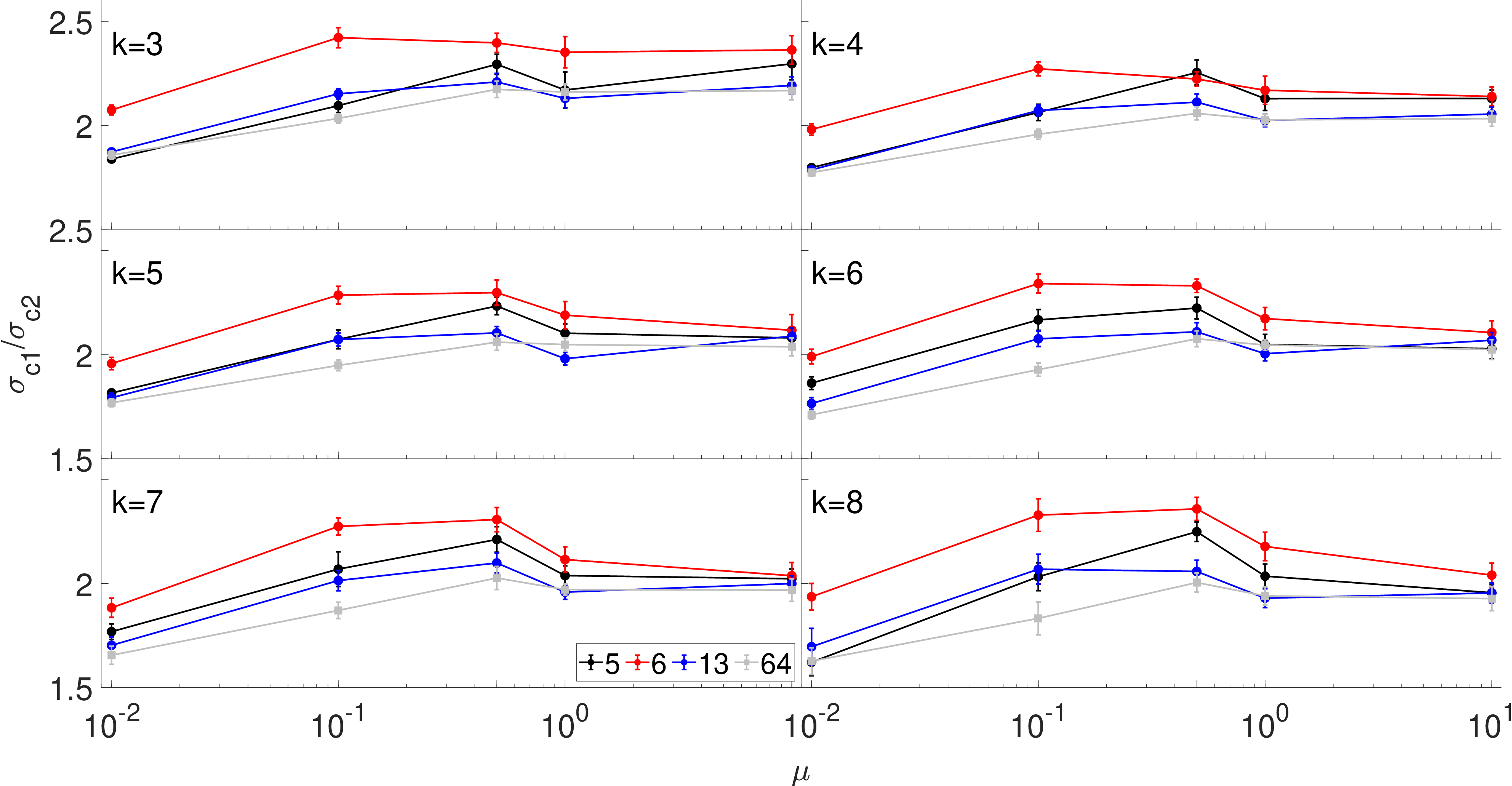}
 \caption{Dependence of the steady state stresses ratio, shown in Fig.\,\ref{fig:cellstress_evolution}  on $\mu$ and $N$.}
 \label{fig:cellstress_CS}%
\end{figure}

The steady-state values of the stress ratio $\sigma_{c1}/\sigma_{c2}$, shown in Fig.~\ref{fig:cellstress_CS}, depend non-monotonically on $\mu$. It increases with $\mu$ up to a maximum between $0.1\leq\mu\leq0.5$ and decreases For $\mu\geq1.0$. The rate of decrease depends on both cell order and particle shape. This is in line with earlier observations~\cite{Trulsson2018, Krengel2025} and can be understood in terms of a competition between inter-particle sliding and rolling which leads to an optimal stability around $\mu=0.5$. We elaborate on this mechanism below. This trend is less pronounced for $N=6$, where the stress ratio is also generally higher than for any other shape. This is because hexagons are more stable, which suppresses this competition.

\begin{figure}[h]
 \includegraphics[width=\columnwidth]{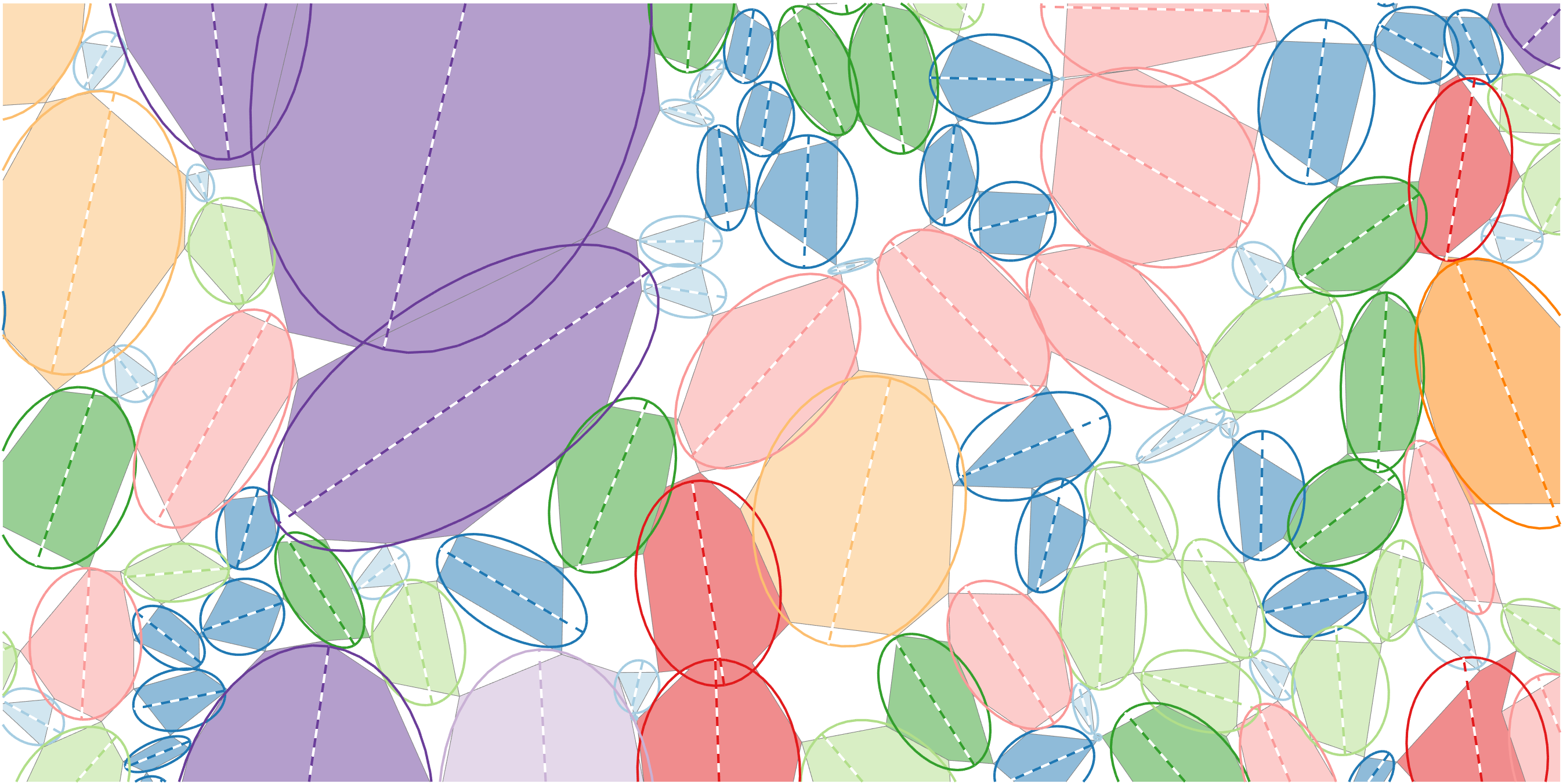}%
 \caption{Example of ellipses fitted to the cells.}%
 \label{fig:ellipse_visualisation}
\end{figure}

\begin{figure}[h]
 \centering
 \includegraphics[width=\columnwidth]{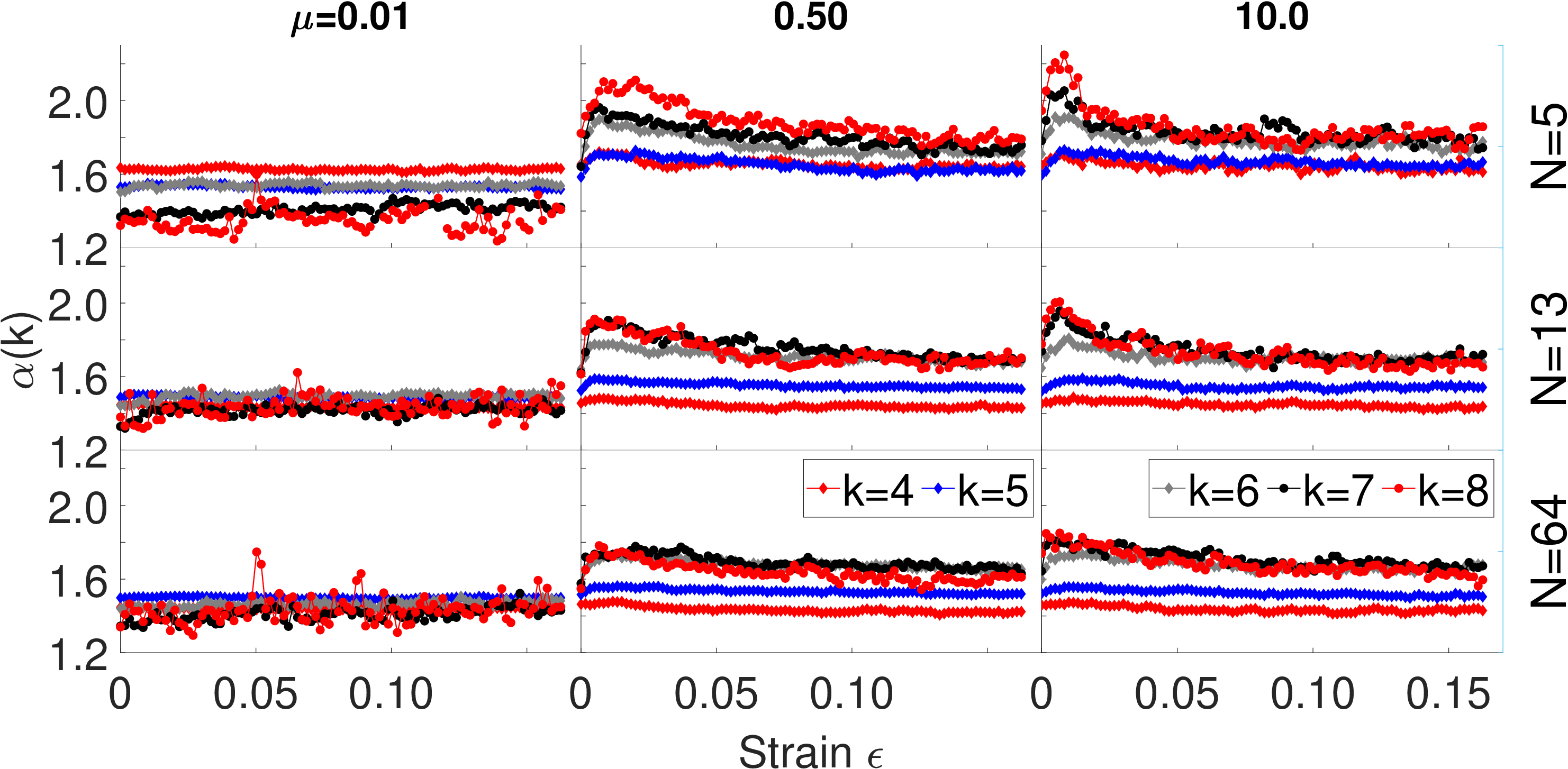}%
 \caption{Evolution of the mean cell aspect ratios. Cells of order $k=3$ are excluded.}
 \label{fig:cellelongation_evo}%
\end{figure}

To investigate cell orientational organisation, we approximate each cell as the ellipse closest to its  contacts~\cite{Jiang2022} (Fig.~\ref{fig:ellipse_visualisation}). We measured the ellipses' aspect ratios, $\alpha$, and major axes orientations, $\theta_c$. We omit in this analysis cells of order $k=3$ for which the ellipse approximation is not useful.
Figure~\ref{fig:cellelongation_evo} shows the evolution of $\alpha$ with strain for combinations of $N$ and $\mu$. At low friction, $\alpha$ is constant regardless of $N$. As $\mu$ increases the mean aspect ratio increases and then decrease to a steady state value. The maximum is more pronounced for large $k$, low $N$, and high $\mu$. It is hardly observable or disappears altogether for $k\leq5$. In this regime, $\alpha$ decreases with increasing cell order. This can be understood in the context of the entropy-mechanical stability competition~\cite{Sun2020}, with the latter dominating at high-order cells. 

\begin{figure}[h]
 \centering
 \includegraphics[width=\columnwidth]{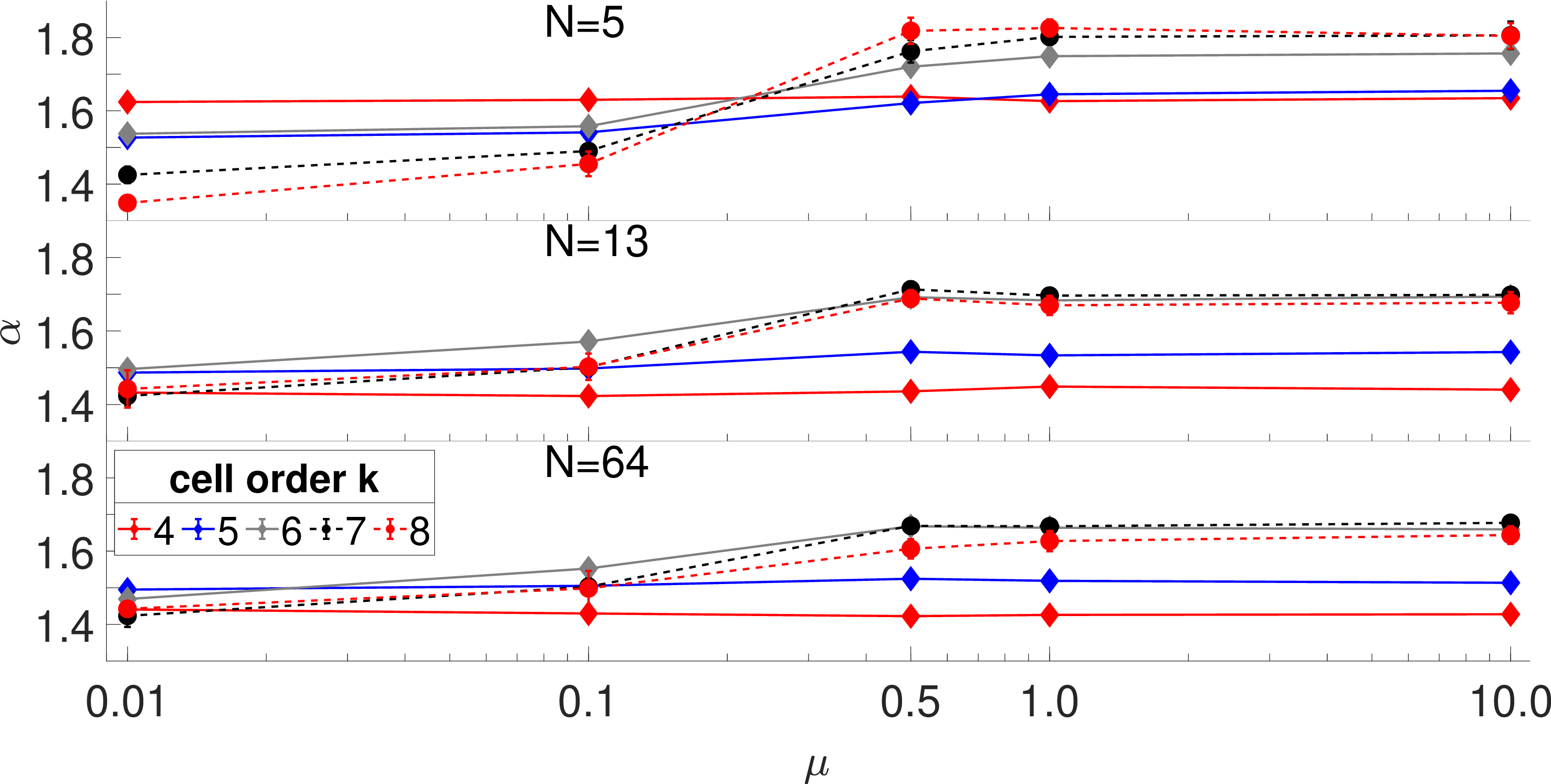}%
 \caption{Cell aspect ratios at steady state as a function of $\mu$ for different angularity $N$.}%
 \label{fig:cellelongation_CS}%
\end{figure}

As expected, the steady-state value of $\alpha$ increases with $\mu$ for all $k>4$ up to a value of $\mu$ in the range $[0.5,1.0]$, which depends on the cell order. The rate of increase itself decreases with both $N$ and $k$ (Fig.~\ref{fig:cellelongation_CS}). Higher friction enhances cells stability allowing more elongated cells to survive shear, an effect that is more pronounced for $k>4$. High angularity also increases elongated cell stability owing to the hindered rearrangement of particles by rotation.

\begin{figure}[h]
 \centering
 \includegraphics[width=\columnwidth]{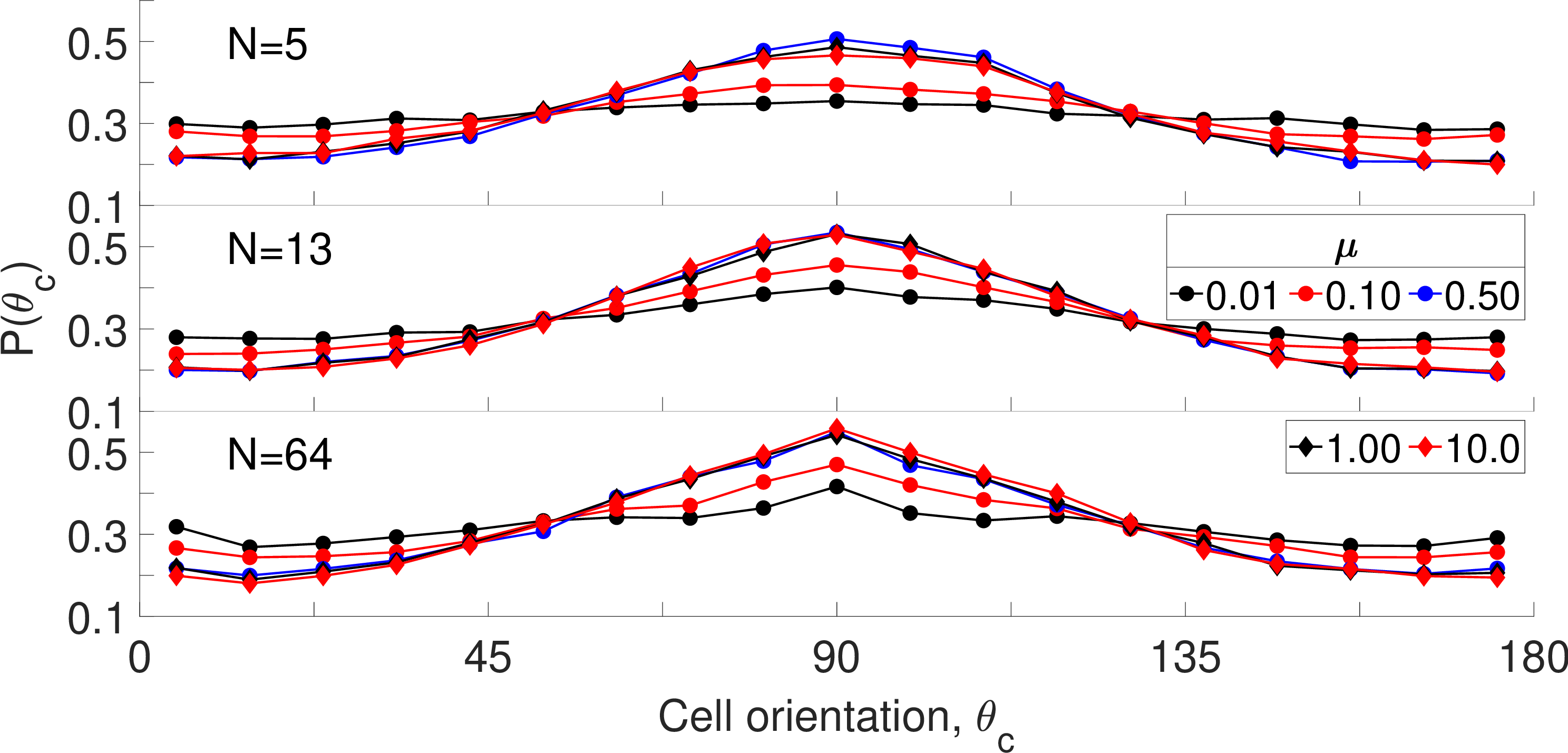}%
 \caption{Overall mean cell orientation, $\theta_{c}$, relative to the horizontal direction, in the steady state.}%
 \label{fig:cellorientation_global}%
\end{figure}

\begin{figure}[h]
 \centering
 \includegraphics[width=\columnwidth]{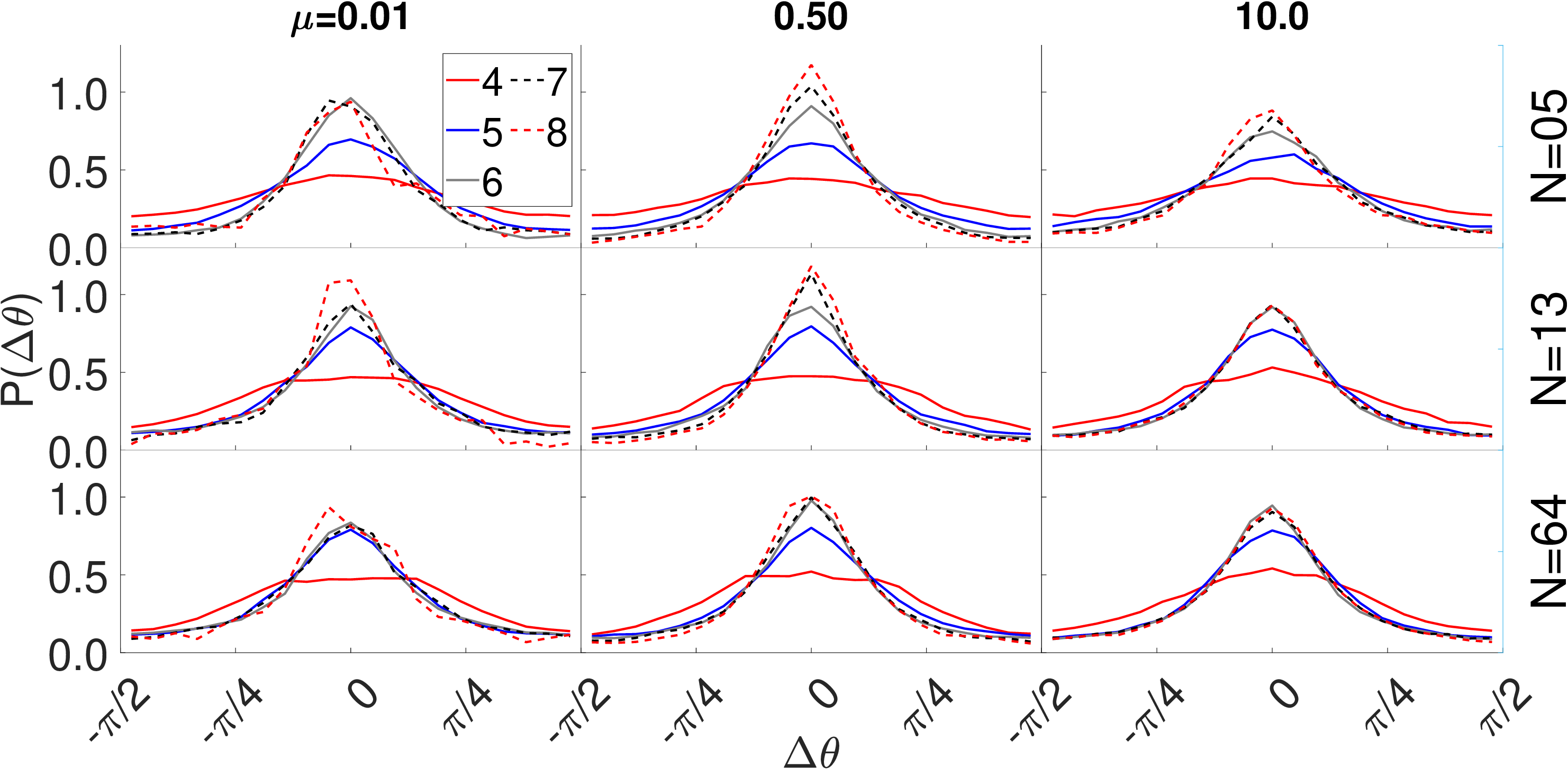}%
 \caption{PDF of $\Delta\theta\equiv\theta_c-\theta_s$ for different cell orders, $k$.}%
 \label{fig:PDF_thetaC_thetaS}%
\end{figure}

Cell organisation is driven by stresses and we find that, for all $N$ and $\mu$, there is a strong correlation between cell orientations $\theta_{c}$ and the local major stress principal axis $\theta_{s}$ (see Fig.~\ref{fig:cellorientation_global}). As Fig.~\ref{fig:PDF_thetaC_thetaS} shows, the correlation becomes weaker as $\mu$ decreases. This is both because cells are less elongated and because the media around cells are more mobile and allow easier cell reorientation.

In Fig.~\ref{fig:PDF_thetaC_thetaS} we plot the distribution of the local difference between the two orientations, $\Delta\theta\equiv\theta_c-\theta_s$, establishing beyond doubt their tendency to align. This phenomenon, which is more pronounced at high-order cells, is as in disc systems~\cite{Matsushima2021,Jiang2022}, suggesting that it is independent of particle angularity. 

\begin{figure}[h]
 \centering
 \includegraphics[width=\columnwidth]{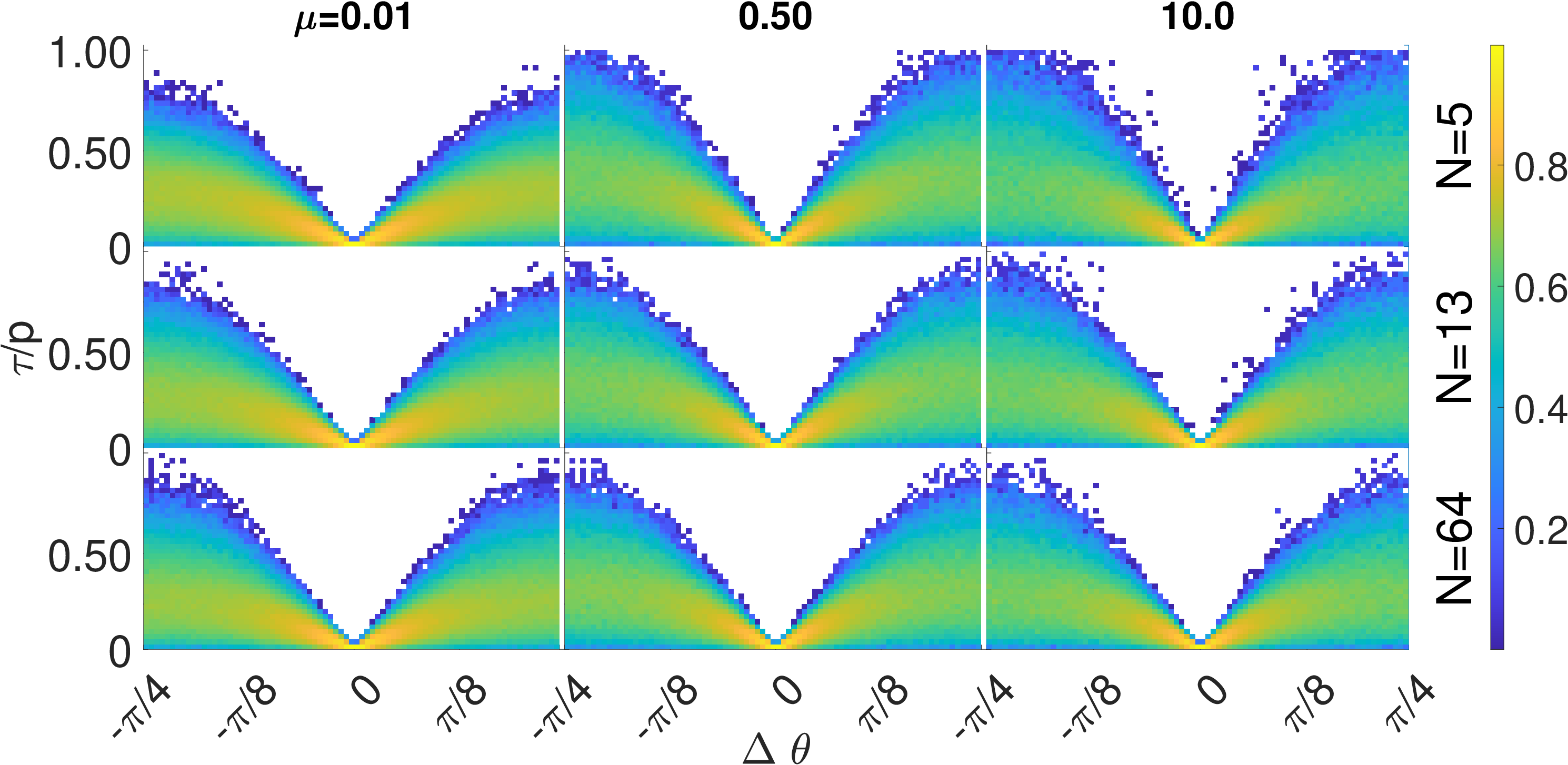}%
 \caption{A scatter plot heat map of cells in the plane spanned by $\Delta\theta$ and $\tau/p$ plane. The colour denotes the relative frequency of cells with given values of $\Delta\theta$ and $\tau/p$. }
 \label{fig:cellorientation_CS}%
\end{figure}

We understand the dynamics that lead to such organisation as follows: Although the process is quasistatic, reorientation of cells requires unbalanced forces to move particles. This means that the cell stresses are momentarily asymmetric, giving rise to small torques that rotate them. The less stable the cell the larger the rotation. 
To quantify a cell's stability, let the axis of its $\sigma_{c1}$ be at an angle $\Delta\theta$ to its major ellipse axis. In the frame of reference aligned with the latter, the cell stress is
\begin{equation}
  \bm{\Sigma} =  \begin{bmatrix}
    \sigma_{c1}+\sigma_{c2}\tan^2\Delta\theta & 
        2q\tan\Delta\theta \\
        2q\tan\Delta\theta & 
        \sigma_{c1}\tan^2\Delta\theta+\sigma_{c2}
    \end{bmatrix}\cos^2\Delta\theta \ .
\label{Stress1}
\end{equation}
A common measure of its stability is the ratio of the off-diagonal stress component to the trace, in this case,
$\tau = \left(\sigma_{c1} -\sigma_{c2}\right)\sin{\left(2\Delta\theta\right)}/\left[2\left(\sigma_{c1} + \sigma_{c2}\right)\right]$. The lower $\tau$ is the more stable the cell. During the dynamic process, the cell is nudged towards higher stability, which amounts to reducing $\Delta\theta$. It is this nudging that gives rise to the observed peak around $\Delta\theta=0$. This can also be seen in the scatter plot of cells in the plane spanned by $\Delta\theta$ and $\tau/p$, shown in Fig.~\ref{fig:cellorientation_CS}.

Generally, there is a competition between two cell rearrangement mechanism~\cite{Krengel2025}. Particles contacting along edges can slide relative to one another while corner-edge and (the rarer) corner-corner contacts can change mainly by a relative rotation. For a given friction coefficient, the more angular the particles the easier it is to rearrange by rotation. Increasing $\mu$ for systems of the same $N$, would also favour rotation over sliding. However, as $\mu\to0$ for a given $N$, the rotation-vs-sliding competition goes from being determined by the two particles in contact to depending on the complex many-body dynamics of their environment, which becomes more fluid.  It is the interplay between these mechanisms that leads to the difference between the behaviour of the $N=5$ particles and that of high-$N$ and discs.

\section{Conclusions}\label{sec:conclusions}
To conclude, we have studied the effects of particle angularity on self-organisation in quasi-static dynamics of 2D granular assemblies. We simulated regular polygonal particles with gradually sharper corners to test which of the indicators observed in disc systems survive increasing angularity. 
We have found that some indicators are independent of angularity while others are not. Specifically, as in disc systems, individual cell stresses are well aligned with their orientations. Similarly, the PDFs of the scaled cell stress ratios, $P\left(\hat{h}=h/\bar{h}_k\right)$, collapse to a master Weibull form for all cell orders $k$. 
However, angularity affects the sensitivity of these PDFs to friction; as $N$ decreases, the collapsed $P\left(\hat{h}\right)$ appear to shift systematically toward lower values of $\hat{h}$ as $\mu$ decreases. While one expects cells in low friction assemblies to be less stable, the reason for this difference is unclear. Since the angular corners are expected to have similar traction as in discs, we conjecture that the difference must stem from increased sensitivity to sliding of contacts between flat surfaces. It would be interesting to test this conjecture in future studies.

Another intriguing finding is that the steady-state behaviour changes around $\mu=0.5$: the approach to steady state is slower (Fig.\,\ref{fig:cellstress_CS}), there appears a peak in the mean ratio $\sigma_{c1}/\sigma_{c2}$, and the resistance to shear increases. These phenomena become more noticeable with increasing angularity. These observations suggest an intricate competition between the sliding and rolling mechanisms that govern re-organisation. The probability of surface-to-surface contacts increases with angularity, increasing the sliding probability but reduces the rolling probability. As $\mu$ increases, the probability of rolling increases, but the barrier to rolling is higher for high-angularity particles. Nevertheless, once rolling takes place, the displacements of centres of high-angularity particles are larger. This competition gives rise to optimal stability at an intermediate value of $\mu$,which happens to be around $0.5$ in our bi-disperse polygonal systems.

In principle, our analysis can be extended to three dimensions (3D). However, to this end some hurdles need to be overcome. While contact data can be readily generated and algorithms to identify 3D cells exist~\cite{Chueire2023}, classification of cells into families is significantly more complex. Cells are polyhedra that are characterized by a number of parameters, as opposed to the one order parameter in two dimensions. Each cell $c$ is surrounded by two types of polygonal faces, $N^c_p$, which are shared with the particles surrounding it, and $N^c_t$, which constitute throats to nearest-neighbour cells. To characterise cell $c$, we then need to specify: $N^c_p$; $N^c_t$; the specific realisation of the order distribution of each of the $N^c_p$ faces, $P_p\left(k\right)$  ($k=1, 2, …, N^c_p $); and the specific realisation of the order distribution of the $N^c_t$ throat faces, $P_t\left(n\right)$ ($n=1, 2, \dots, N^c_t$). A family of cells family not only has the same $N^c_p$ and $N^c_t$ faces but it also has the exact same realisation of orders in each type. Given that cells could have 12-14 faces, the number of different combinations of having particle and throat faces is larger than 7500. Even assuming that the orders are distributed between 3 and 9 (the average must be 6), the number of face order realisations for each combination of these is ${\mathcal{O}\left( 10^{5} \right)}$.  Moreover, for good statistics, one would need to have $10^2-10^3$ specimens of each family member, which means running $10^7-10^8$ different systems. This is not feasible with current resources. 

\section*{CRediT authorship contribution statement}
\textbf{DK:}  Writing – original draft, Writing – review \& editing, Visualization, Validation, Software, Methodology, Investigation, Formal analysis, Data curation, Conceptualization. 
\textbf{HJ:} Writing – review \& editing, Software, Visualization, Methodology. 
\textbf{RB:}: Writing – original draft, Writing – review \& editing, Data curation, Conceptualization, Supervision
\textbf{TM:} Conceptualization, Supervision, Resources.

\begin{acknowledgments}
DK, TM, and RB are grateful for the support of the Grant-in-Aid for Scientific Research 21KK0071 from the Japan Society for the Promotion of Science (JSPS). DK and TM  are also grateful for the support of the Grant-in-Aid for Scientific Research 21H01422 from JSPS.
\end{acknowledgments}

\appendix
\section{Polygonal DEM method}
The polygonal DEM used in this work is based on the monograph by Matuttis and Chen\,\cite{Matuttis2014}. The interaction force between polygons is based on the particle overlap area $A$ as a stand-in for the physical deformation of two particles (P1 and P2) during contact. We use the centroid $\mathbf{P}$ of the overlap area as force point. The intersection points define the tangential direction $\bm{t}$, which also fixes the normal direction of the contact, $\mathbf{n}$. The elastic normal force
\begin{equation}
 F_{\mathrm{el}} = \frac{YA}{l},
\end{equation}
with the Youngs modulus $Y$ and a characteristic length
\begin{equation}
 l=\frac{4|\mathbf{r}_{\mathrm{A}}||\mathbf{r}_{\mathrm{B}}|}{|\mathbf{r}_{\mathrm{A}}|+|\mathbf{r}_{\mathrm{B}}|}
\end{equation}
with the contact vectors $\mathbf{r}_{\mathrm{P1,P2}}$ from the the centres of mass to $\mathbf{P}$. We further have a dissipative normal force
\begin{equation}
 F_{\mathrm{diss}} = \gamma\sqrt{mY}\frac{\dot{A}}{l},
\end{equation}
with the reduced mass $1/m = 1/m_{\mathrm{A}}+1/m_{\mathrm{B}}$ and a damping constant $\gamma$. The total normal force is
\begin{equation}
 F_{\mathrm{N}} = F_{\mathrm{el}}+F_{\mathrm{diss}}.
\end{equation}
Friction acting in tangential direction is based on the Cundall-Strack model\,\cite{Cundall1979},
\begin{equation}
 F_{\mathrm{T}}(t)=\left\{ \begin{array}{ll}
  F_{\mathrm{T}}(t-\Delta t)-k_{\mathrm{T}}\Delta t v_{\mathrm{T}}, & \lvert F_{\mathrm{T}}(t)\rvert \leq \mu F_{\mathrm{N}} \\
	\mathrm{sgn}(F_{\mathrm{T}}(t-\Delta t))\mu F_{\mathrm{N}}, & \lvert F_{\mathrm{T}}(t)\rvert > \mu F_{\mathrm{N}}
 \end{array}\right. ,
 \label{eq:CSfriction}
\end{equation}
 with the relative tangential velocity $v_{\mathrm{T}}$, the tangential stiffness of a spring $k_{\mathrm{T}}$, and $\mu$ as the inter-particle friction coefficient. The total force $\mathbf{F} = F_{\mathrm{N}}\mathbf{n} +F_{\mathrm{T}}\mathbf{t}$ then induces a torque on the particles,
\begin{equation}
 \tau_{P1,P2} = \mathbf{r}_{\mathrm{P1,P2}}\times \mathbf{F}.
\end{equation}

\bibliography{polygon_cells_v2.bib}
\end{document}